# Observation of molecular and polymeric nitrogen stuffed NaCl ionic layers

Ping Ning[1,5], Yifan Tian[1,5], Guangtao Liu[1,2,5], Hongbo Wang[2], Qingyang Hu[3,*], Hanyu Liu[1,2,4,*], Mi Zhou[1,2,*],

Yanming Ma[1,2,4,*]

**Abstract**

Sodium chloride (NaCl), a ubiquitous and chemically stable compound, has been considered inert under ambient conditions. Its typical B1 structure is highly isotropic without preferential direction, favoring the growth of a three-dimensional network of strong Na-Cl ionic bonds. Here, we employ first-principles structural searching and synchrotron X-ray diffraction to unravel an unexpected chemical reaction between NaCl and $N_2$ to produce a hybrid salt-$NaCl(N_2)_2$, where $N_2$ molecules break the isotropic NaCl structure into two-dimensional layers upon synthesis at 50 GPa. In contrast to the insulating properties of pristine NaCl, the electronic bandgap of the $N_2$-stuffed NaCl narrowed to 1.8 eV, becoming an indirect bandgap semiconductor. Further compression to 130 GPa induced the polymerization of N atoms into zigzag N-chains. Our findings not only demonstrate the possibility of unusual N-chemistry under extreme conditions, but also suggest a feasible approach for the design of layered NaCl frameworks to modulate the polymerization of nitrogen.

[1]Key Laboratory of Material Simulation Methods and Software of Ministry of Education, College of Physics, Jilin University, Changchun 130012, China. [2]State Key Laboratory of Superhard Materials, College of Physics, Jilin University, Changchun 130012, China. [3]Center for High Pressure Science and Technology Advanced Research, Beijing, 100094, China. [4]International Center of Future Science, Jilin University, Changchun 130012, China. [5]These authors contributed equally: Ping Ning, Yifan Tian, Guangtao Liu

*E-mail: qingyang.hu@hpstar.ac.cn; hanyuliu@jlu.edu.cn; mzhou@jlu.edu.cn; mym@jlu.edu.cn;

## Introduction

NaCl is the most abundant salt in seawater and an essential nutrient for life-beings. Its chemical properties have been extensively studied under both atmospheric and high-pressure conditions. Due to its non-reactivity with major components of the atmosphere, sea salt is readily extracted from seawater by evaporation and crystallization. Under ambient conditions, NaCl does not participate in oxidation or reduction reactions in general due to the strong electrostatic forces between its ions and the stability of the $Na^+$ and $Cl^-$ states. Only exposing under extreme acidic environments or highly oxidizing agents, NaCl may dissociate to form stronger ionic compounds[1-4]. Besides, applying pressure modulates the chemical properties by shortening the Na-Cl bond length, and previous high-pressure experiments have revealed redox chemistry between NaCl and reactive gases. For instance, applying tens of gigapascal (GPa) on the mixture of NaCl and $Cl_2$ stabilize non-stoichiometric Na-Cl compounds such as $Na_3Cl$ and $NaCl_3$[5].

On the atomistic scale, the chemical properties of NaCl have profound relation with its crystal structure. The rock salt structure (B1), named after NaCl, is an archetypal structural model in the textbook of crystallography and chemistry, serving as a fundamental temple for understanding the crystal structures of many other ionic compounds. Each $Na^+$ or $Cl^-$ is coordinated with six ions of the opposite charge. Due to the cubic symmetry, the elastic and electronic properties of NaCl are isotropic. This structural isotropy is predicted to persist in NaCl to megabar pressures, even under pressures where the strengths of the Na-Cl ionic bonds become weakened due to the decrease of electronegativity and chemical hardness of sodium and chlorine[6]. For instance, NaCl undergoes a typical phase transition to a body-centered cubic structure (eight-coordination, CsCl-B2 phase), which is stable at least to 304 GPa[7,8]. It was theoretically predicted that NaCl loses its structural isotropy at above 322 GPa, in which the B2 phase transforms to a sequence of structurally anisotropic phases like oC8-type layers, oI8-type chains and oP16-type tunnel at 322 GPa, 645 GPa and 683 GPa, respectively[9].

Other than pressurization, the structural isotropy of NaCl may be interrupted by introducing covalently bonded molecules into its crystal lattice. It was recently predicted that small molecules (sM) like $H_2$, $N_2$, $H_2O$, $NH_3$, $CH_4$, and $C_2H_6$, can insert into the interstitial sites and distort the cubic structure, forming hybrid salts[10,11]. The recent synthesis of $NaClH_2$ unravel the unconventional chemistry to create hybrid salt under high pressure, but $H_2$ is rather small in size and failed to broke the isotropy of the pristine NaCl[12]. On this basis, we combine *in situ* x-ray diffraction (XRD) and first-principal simulations to trigger high-pressure chemistry between NaCl and $N_2$ to synthesize an unexpected layered $NaCl(N_2)_2$ structure with unique electronic structures and N-packing patterns.

## Results and discussion
### Experimental synthesis of $NaCl(N_2)_2$

High-purity NaCl powders were loaded with $N_2$ as pressure medium in laser-heated diamond anvil cells. These samples were compressed to 40-55 GPa at ambient temperature and then laser heated to approximately 2000 K using an yttrium-aluminum-garnet laser (1064 nm), and then the samples were compressed 130-140 GPa at room temperature. We observed the color of NaCl samples changed from transparent to translucent in spots after shining laser (Fig. 1(a)), suggesting the occurrence of chemical reaction. Raman spectroscopy and synchrotron XRD measurements were then carried out to investigate the crystal structure of the product.

In this experiment series, Raman spectroscopy[13] is mainly used to characterize N-related chemical bonds, and this technique is well established to distinguished sophisticated N-structure such as chains[14,15], rings[16-19], and layers[20,21]. As shown in Fig. 1(b), black and red curves compare the Raman spectra before and after laser heating. We noticed several broad bands located in the low-frequency region (100-500 cm$^{-1}$), which were previously identified as the lattice vibrations of $N_2$[22]. Sharp and intense peaks were observed in the vicinity of 2400 cm$^{-1}$, and they can be attributed to the stretching modes of N-N triple bond[22]. We highlight the newly excited Raman peaks after shining laser (the red curve in Fig. 1(b)), which positioned at 2321.6 cm$^{-1}$ at 40 GPa. According to group theory, either the B1 or B2 phases of NaCl are Raman inactive. Therefore, this Raman peaks should stem from the product of high-pressure chemistry between $N_2$ and NaCl. Its proximity to N≡N stretching mode suggests the onset of softened N≡N bonds which have longer bond lengths than those in pure nitrogen. Such bond softening was also recognized in N-stuffed Re nitride (ReN$_8$·$x$N$_2$)[23]. Meanwhile, peaks in the low-frequency region become sharper, indicating crystal grain growth and the formation of a more ordered crystal structure. To determine its stability region, Raman spectra were also collected during the decompression process (Fig. 1(c)). We should note that the slope of frequency for the softened N≡N bond (the red dots in Fig. 1(c)) is lower than that of pure nitrogen (the black dots in Fig. 1(c)), indicating that the N≡N bonds in NaCl-$N_2$ compound exhibit lower sensitivity to the external compression, owing to the effects of the synergistic combination of NaCl and nitrogen components.

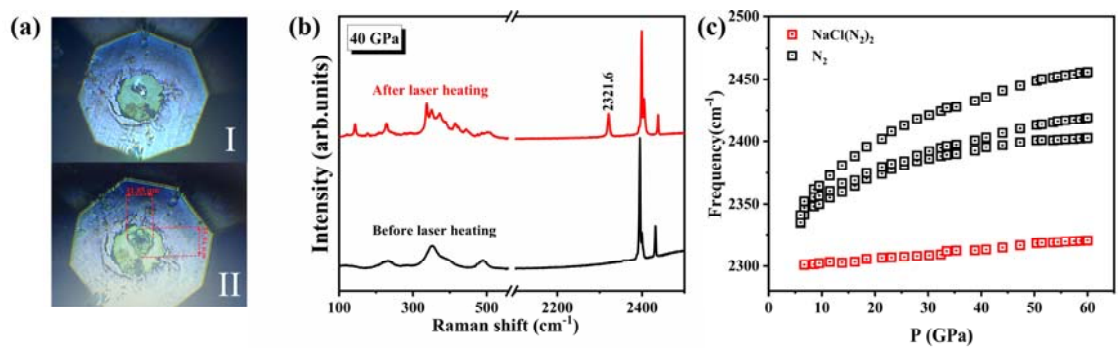

**Fig. 1.** Characterization of NaCl(N$_2$)$_2$ using Raman spectroscopy. (a) Optical images of the sample chamber before (inset Ⅰ) and after laser heating (inset Ⅱ). (b) Raman spectra of the NaCl-N$_2$ compound before and after laser-heating. (c) Raman frequency-pressure curves of the NaCl-N$_2$ compound.

**The crystal structure of NaCl(N$_2$)$_2$**

Before presenting XRD data, we performed structure-searching simulations for NaCl-N$_2$ system to provide candidate high-pressure structures. This calculation was conducted within the vdW-DF2 density functional[24] with emphasis on the description of long-range interactions. This treatment is motivated by the recent experiment[12] implying the van der Waals (vdW) effects between sM and NaCl have played key roles in predicting their crystal structures. Specifically, our calculations were initiated with various NaCl-N$_2$ ratios and up to 4 formula units (f.u.) per cell. Fig. 2(b) is the convex-hull curve of NaCl-N$_2$ compounds at 50 GPa and the stable structure of NaCl(N$_2$)$_2$ is labelled by the red star Fig. 2(c). It adopts an orthorhombic lattice with the symmetry space group of *Ccca*, which could be viewed as an intercalation of N$_2$ molecules into the layered NaCl with planar tetracoordinate Na and Cl atoms. To the best of our knowledge, it is rare to stabilize ionic two-dimensional (2D) structure at such high pressure. It is interestingly seen that similar theoretical structures containing intercalated N$_2$ were also predicted in a recent report[11].

Structures obtained from simulation were then used in interpreting XRD patterns from synchrotron-based experiment. The emerging peaks located at 5.21°, 11.28° and 11.40° (Fig. 2(a)) match well with the (200), (220) and (112) crystal planes of the predicted stable *Ccca*-type NaCl(N$_2$)$_2$. Lattice parameters at 55 GPa is further refined to $a$ = 9.10(6) Å, $b$ = 4.75(7) Å, $c$ = 4.80(0) Å ($V$ = 207.92 Å$^3$) that agree within reasonable uncertainties with calculation of NaCl(N$_2$)$_2$ (Fig. 2(d)). For the newly synthesized structure, in each NaCl layer, the bond lengths between Na and Cl atoms are 2.38 and 2.40 Å respectively, which are slightly shorter than those in the *Pm*-3*m* NaCl (2.51 Å at 55 GPa). However, the N≡N bond length of NaCl(N$_2$)$_2$ is measured as 1.10 Å, slightly longer than that in pure N$_2$ (1.08 Å) at the comparable pressure, echoing the softened stretching Raman mode in Fig.1 (b). The 2D structure is defined by the distance between adjacent NaCl layers, *e.g.* 4.83 Å at 50 GPa, which is much longer than the ionic bond length in NaCl. This substantial interlayer distance suggests that the ionic interactions between adjacent NaCl layers have been destroyed. Unlike the previously synthesized NaClH$_2$, where H$_2$ molecular units occupy the interstitial sites of the NaCl lattice[12], N$_2$ molecular units are inserted in between deionized NaCl layers and transform NaCl into 2D structure. Under ambient conditions, similar layered NaCl could only be epitaxially grown on metallic substrates such as Fe(001) and Cu(111) under ultra-high vacuum conditions[25,26].

The elastic parameters for NaCl(N$_2$)$_2$ were then fitted by a third-order Birch-Murnaghan equation of state (EOS) and fixing $B_0$' = 4.0, where $B_0$ = 22.4(3) GPa and $V_0$ = 103.2(0) Å$^3$/f.u. In comparison, theoretical calculation predicted $B_0$ = 28.6(2) GPa and $V_0$ = 98.99(1) Å$^3$/f.u. (Fig. 2(d)), indicating this material is rather much less compressible than NaCl ($B_0$ = 39.6(0) GPa). We should note that elastic anisotropy has become prominent through the NaCl-N$_2$ high-pressure chemistry, as the *a* axis if much compressible than *b* and *c* axes. Such anisotropy paves the way of novel mechanochemistry by converting hydrostatic pressure to preferential stress on its layers, which will be illustrated in the following sessions.

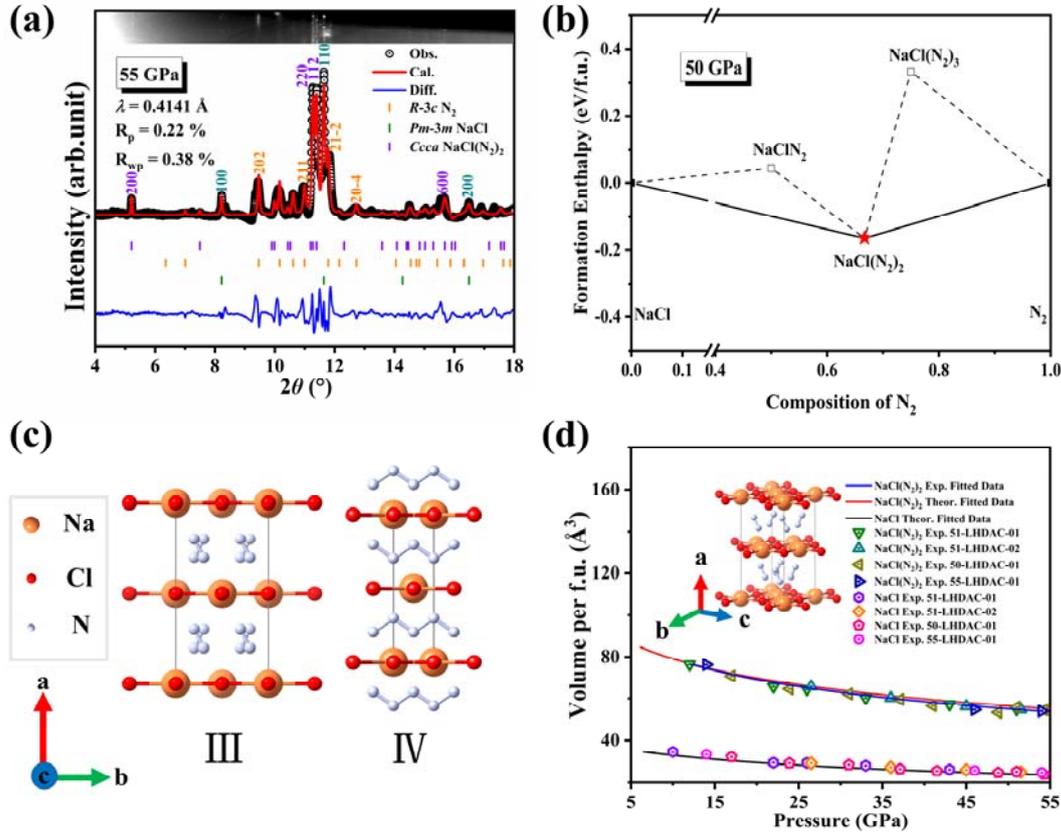

**Fig. 2.** Characterizing NaCl(N$_2$)$_2$ structure through XRD. (a) The synchrotron X-ray diffraction pattern of the NaCl-N$_2$ compound from 55-LHDAC-02 was obtained and the Rietveld refinement of the *Ccca*-NaCl(N$_2$)$_2$ structure at 55 GPa. The cake view of the collected XRD pattern is shown in the upper inset. The black open circles, red solid curve, and blue solid curve represent observed data, the Rietveld fit, and the residual intensity, respectively. (b) The formation enthalpy as a function of x for (NaCl)$_{1-x}$N$_x$ at various pressures. The red star highlights NaCl(N$_2$)$_2$ as a thermodynamically stable phase. (c) The predicted crystal structures of NaCl (N$_2$)$_2$ at 50 GPa (inset Ⅲ) and NaClN$_4$ at 130 GPa (inset Ⅳ), respectively. Atoms of sodium (Na), chlorine (Cl), and nitrogen (N) are depicted in orange, red, and gray, respectively. (d) Volumes as a function of pressure for NaCl (N$_2$)$_2$ and NaCl. The triangular symbols, blue solid curve, and red solid curve represent the experimental data, fit, and simulation for NaCl (N$_2$)$_2$, respectively. The various signs below and solid black curve represent the experimental data and simulation for NaCl, respectively.

## NaCl(N$_2$)$_2$: a narrow bandgap semiconductor

The electronic structures of NaCl(N$_2$)$_2$ hinge on its unique atomic packing pattern. For metals and ionic solids, pressurization typically increases the coordination number by densifying atomic packing[27-29]. The archetypal B1-B2 transition in NaCl promotes the coordination from 6 to 8[7]. In stark contrast, the synthesis of 2D NaCl(N$_2$)$_2$ reduced the coordination number of Na from 8 (in the B2-phase) to 4, resulting in the formation of layered NaCl with a planar tetracoordinate structure. In order to decipher the nature of chemical bonding in NaCl(N$_2$)$_2$, the electron localization functions (ELFs)[30] were calculated and shown in Fig. 3(c). At 50 GPa and static conditions, the chemical bonds between Na-Cl are strongly polarized, suggesting the ionic nature of NaCl solid layers. In contrast, high ELF values are observed between the N-N atoms, indicating highly localized electrons and the covalent bonding of N$_2$ molecular units in NaCl(N$_2$)$_2$. The NaCl(N$_2$)$_2$ is a typical indirect band-gap semi-conductor with a band gap width of 1.8 eV at 50 GPa, which falls into the visible light regime. According to its stability range, the band gap width increases to 3.0 eV by releasing pressure to 7 GPa. This pressure tunability makes it to cover wide light spectrum from infrared red to visible and possible implicate to optical devises through the technique called gimmickry, which can literate high-pressure properties by encapsulation for manufacturing larger-sized devices[31].

We further plot the band structures[32] in Fig. 3(a), in which the radii of colored circles proportional to the weights of the corresponding elements. The introduction of $N_2$ molecules reduces the band gap, as the conduction bands are primarily contributed by the N atoms. As shown in Fig. 3(a), the bands at CBM (conduction band minimum) and VBM (valance band maximum) of $NaCl(N_2)_2$ are flat. Consequently, the hole effective mass ($m_h^*$)[33] was fitted along the $k$-point path from Γ to Z with $m_h^*$ = -2.273 $m_0$, and the electron effective mass ($m_e^*$) was fitted along the $k$-point path from S to Y and S to X with $m_e^*$ values of -11.154 $m_0$ and -5.679 $m_0$, respectively. The high effective mass results in low carrier mobility, which may lead to a slower response to electric fields, helping improve stability and reduce noise, making the system less sensitive to external disturbances.

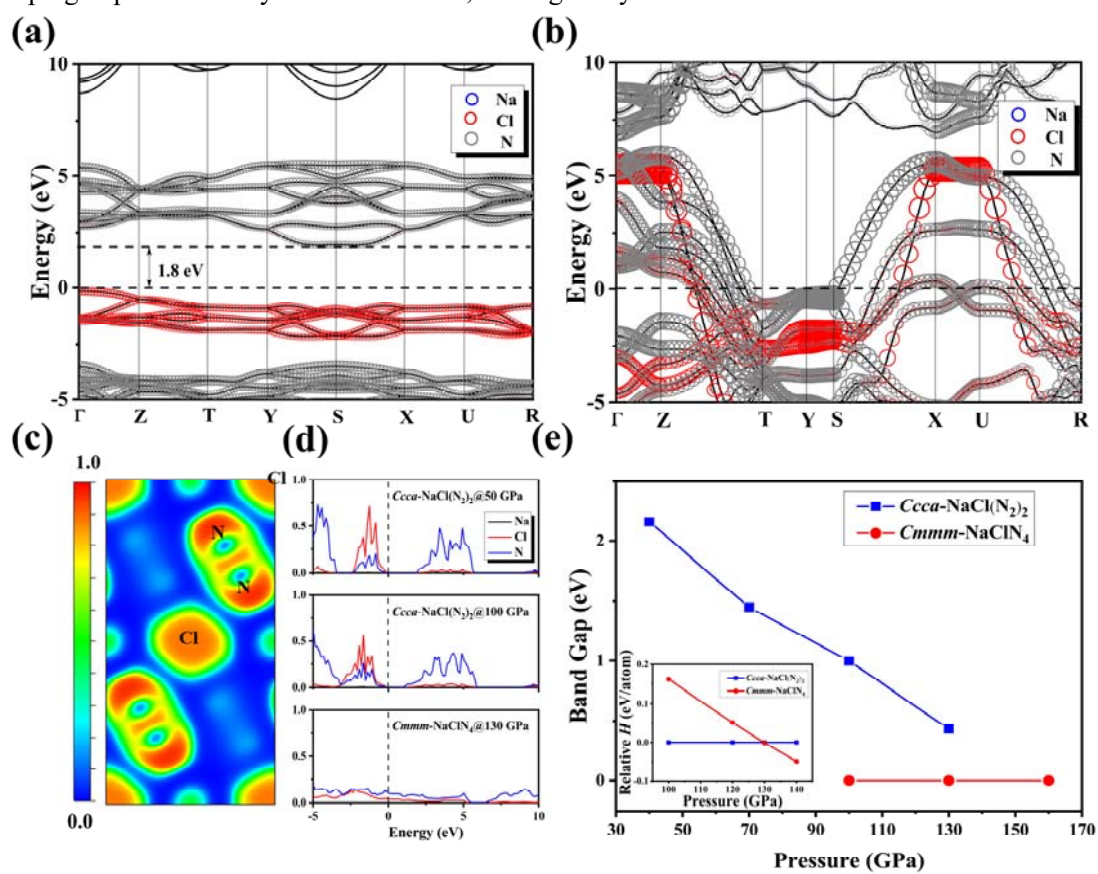

**Fig. 3.** (a) Calculated band structures of $NaCl(N_2)_2$ at 50 GPa (b) Calculated band structures of $NaClN_4$ at 130 GPa (c) Calculated ELFs contour maps plot in (0 0 1) plain of $NaCl(N_2)_2$ at 50 GPa (d) Calculated PDOS of $NaCl(N_2)_2$ and $NaClN_4$ at 50, 100 and 130 GPa, respectively. (d) Pressure dependence of band gaps. The inset shows the pressure dependence of Relative H.

## Mechanochemistry of 2D $NaCl(N_2)_2$ for high-density energy materials

The intriguing crystal and electronic structure of 2D $NaCl(N_2)_2$ motivate us to compress $NaCl(N_2)_2$ beyond its synthesis pressure and we observed the transformation into $NaClN_4$ (space group of *Cmmm*) with polymerized N at 130 GPa. Specifically, the nearest-neighbor N distance of $NaClN_4$ is 1.28 Å [110.30°], shorter than those found in bulk N polymorphs (*e.g.* 1.34-1.44 Å in BP-

N)[20,21,35,36]. Meanwhile, the shortest adjacent interchain distance is 2.19 Å at 130 GPa. Here, mechanical energy is mainly applied to the stuffed $N_2$ molecules and achieve the polymerization of N-molecule by enforcing anisotropic compression on the interlayer spaces. Since critical polymerized pressure of $NaClN_4$ is lower than that of pure $N_2$ phases such as black-phosphorus, α-arsenic, the layered-boat phase and *Cmcm* phase[21,37,38], the mechanochemistry under the framework of NaCl-layer is more efficient in storing energy in N-bonds.

Upon reaching ~130 GPa, the band gap of $NaClN_4$ closes and becomes a metallic phase, as shown in Fig. 3(e). Our theoretical calculation suggest that metallization pressure is even lower than the that of N polymerization. As a result, the projected density of states in Fig. 3(d) near the Fermi energy level are mainly dominated by nitrogen atoms, indicating that the polymerization behavior may have contributed to its metallization. To date, synthesized polymerized nitrogen has been characterized by a three-coordinated N-N bond in various structural forms: a three-dimensional cubic gauge structure[38], 2D layered structures composed of $N_6$ hexagons[20,35] or a puckered layered structure that incorporates both zigzag and armchair configurations[21]. Unlike the three- or two-dimensional polymerization of pure nitrogen, polymerized nitrogen achieved by mechanochemistry in $NaClN_4$ is arranged by long-range ordered zigzag N-chains. The NaCl ionic layers define a chemical framework that forms one-dimensional nitrogen strings in confined spaces.

## Conclusions

In summary, our experiment successfully synthesized the $NaCl(N_2)_2$ at 55 GPa and 2000 K. This hybrid-salt is viewed as an intercalation of $N_2$ molecules into a layered NaCl by forming planar tetracoordinate Na and Cl atoms. The interlayer distance between adjacent NaCl layers is much greater than the typical ionic bond length of pure NaCl, defining its 2D structure. Electronic band structure calculations show that $NaCl(N_2)_2$ is an indirect band-gap semi-conductor with a band gap of 1.8 eV at 50 GPa. In addition, our further results on $NaCl(N_2)_2$ at higher pressures reveal the polymerization of N atoms within the NaCl layers, forming infinite zigzag N-chains at 130 GPa. These results open up a new route for exploring novel prototype structures of salt-sM compounds under HTHP conditions, and showcases a new synthesis methodology for polymathic sM through synergistic combination of high-pressure mechanochemistry as well as chemical pre-compression.

## Acknowledgements

This research was supported by the National Key Research and Development Program of China (Grant No. 2021YFA1400503 and 2021YFA1400203), and the National Natural Science Foundation of China (Grant No. 52288102, 52090024, T2425016, T2225013, 12474011 and U2230401). Fundamental Research Funds for the Central Universities (Jilin University, JLU), Program for JLU Science and Technology Innovative Research Team (2021TD-05), the Fundamental Research Funds for the Central Universities and computing facilities at the High-Performance Computing Centre of Jilin University. The XRD measurements were performed at the BL15U1 at Shanghai Synchrotron Radiation Facility and the BL10XU of SPring-8.